\documentclass[12pt]{article}%
\usepackage{amsmath}%
\usepackage{amsfonts}%
\usepackage{amssymb}%
\usepackage{graphicx}
\providecommand{\U}[1]{\protect\rule{.1in}{.1in}}

\begin{document}

\title{Principle of Maximum Fisher Information from Hardy's Axioms Applied to
Statistical Systems}
\author{B. Roy Frieden\\College of Optical Sciences, \\University of Arizona, \\Tucson, Arizona, 85721
\and Robert A. Gatenby\\Mathematical Oncology and Radiology,\\Moffitt Cancer Center, \\Tampa, Florida 33612}
\maketitle

\begin{abstract}
Consider a finite-sized, multidimensional system in parameter state
\textbf{a}. The system is either at statistical equilibrium or general
non-equilibrium, and may obey either classical or quantum physics. L. Hardy's
mathematical axioms provide a basis for the physics obeyed by\ any such
system. One axiom is that the number $N$ of distinguishable states
\textbf{a}\ in the system obeys $N=\max.$ This assumes that $N$ is known as
deterministic prior knowledge. However, most observed systems suffer
statistical fluctuations, for which $N$ is therefore only known approximately.
Then what happens if the scope of the axiom $N=\max$ is extended to include
such observed systems? It is found that the state \textbf{a} of the system
must obey a principle of maximum Fisher information, $I=I_{\max}$.\ This is
important because many physical laws have been derived, \emph{assuming} as a
working hypothesis that $I=I_{\max}$. These derivations include uses of the
principle of Extreme physical information (EPI). Examples of such derivations
were of the De Broglie wave hypothesis, quantum wave equations, Maxwell's
equations, new laws of biology (e.g. of Coulomb force-directed cell
development, and of in situ cancer growth), and new laws of economic
fluctuation and investment. That the principle $I=I_{\max}$ itself
\emph{derives,} from suitably extended Hardy axioms, thereby eliminates its
need to be \emph{assumed} in these derivations.\ Thus, uses of $I=I_{\max}$
and EPI express physics at its most fundamental level -- its \emph{axiomatic
basis} in math.\medskip

\noindent\textbf{Key Words: }axioms of physics, maximum Fisher information,
Hardy's axioms, extreme physical information

\end{abstract}

\section{ Background:\ On Hardy's Axioms}

It is long known that all classical physics follows the ZFC
(Zermelo-Fraenkel-Choice) axioms of mathematical set theory. \ Quantum
mechanics does not, however, obey these axioms. \ Nevetheless quantum
mechanics does follow from many alternative sets of axioms, e.g. approaches by
von Neumann, Mackey, Jauch and Lande and Hardy [1]. Here we concern ourselves
with the particular set of 5 axioms due to Hardy [1]. For brevity, we do not
discuss all five of the axioms, only those relevent to our analysis. These are
Eq. (2) and Axiom 3 listed below Eq. (6).

The \emph{motivation} for this study is as follows. The probability
$p($\textbf{x}$)$ or amplitude $\Psi($\textbf{x}$)$ laws governing physical
systems have been derived [2-20] on the basis of a unifying hypothesis or
\emph{assumption}: that the Fisher information [21],[22] in each obeys
$I=\max\equiv I_{\max}$, subject to a physical constraint. \ The derivations
often take the form of a principle of Extreme physical information (EPI), as
discussed below. \ Can the assumption $I=I_{\max}$ be justified?

One of Hardy's axioms is Eq. (2), that the number $N$ of distinguishable
states of the system obeys $N=\max.$However, this value of $N$ is assumed as
deterministically known prior knowledge. But most observed systems suffer
random fluctuations, so that any \emph{observed} value of $N$ is intrinsically
random and subject to error. On the grounds that acknowledging such random
error might lead to a method of determining its probability law, we assume that:

(1) Hardy's axiom $N=\max$ may be applied to \emph{observed, }noise-prone systems.

(2) This observable value of $N$ obeys the Hardy subsystem separation property
$N=N_{A}N_{B},$ as ennumerated in Sec. I B below. \ 

We show that under these assumptions the state \textbf{a} of the system has
the property that its Fisher $I=I_{\max}.$ This, then, will justify past uses
of the latter principle and EPI to derive physical laws governing the observed
system.\ In a nutshell, the aim of this paper is to provide axiomatic
justification for such past derivations of physical laws.

\subsection{Estimating a system state: distinction from the Hardy scenario}

The whole derivation devolves upon determining the number $N$ of
distinguishable states of the system. To determine $N$\ requires knowledge of
\emph{how well} the parameters defining the state may be distinguished. This
in turn asks how well they may be estimated. Hence the problem becomes one of
parameter estimation. Consider, then, an $n$-dimensional system characterized
by a vector (in quantum systems, a Hilbert-) state \textbf{a }$\equiv a_{j},$
$j=1,...,n.$\ Each $a_{j}$ is a dimensional component or 'channel,' $j$ of the
fixed state\ \textbf{a. }State \textbf{a} can be, e.g., that of boson
polarization, or fermion spin, or position, depending on application.\ Each
$a_{j}$ is thereby a degree of freedom of the fixed state \textbf{a}. Each
$a_{j}$, as a trigonometric projection, obeys $0\leq a_{j}\leq l$ \emph{on the
continuum} within an $n$ dimensional cube with a common side $l$ in each
dimension $j$.

It is important to distinguish our system scenario from that in the
introductions to Refs. [1].\ \ There, as an idealized example, $a_{j}$ is the
spin in \emph{a binary problem} where the particle can either have spin up or
spin down. \ In that case there are straightforwardly $N=2$ distinguishable
states of the system. This is known as prior knowledge of the observed
phenomenon. We instead take the experimentalist's view where only their
continuously variable projections are known as data. Such continuity is
required because we shall be using a Fisher-information approach to parameter
estimation. This\emph{ requires }the parameters to be differentiable and,
hence, continuous.

\subsubsection{Random nature of system}

Our major distinctions from the Hardy scenario result from allowing particle
state \textbf{a} to realistically obey random fluctuations \textbf{x}$_{j}$.
These, moreover, obey an unknown probability amplitude law. For example, in
quantum mechanics, the amplitude law governing positional fluctuations
\textbf{x}$_{j}$ of a nonrelativistic particle from a state position
\textbf{a} obeys \emph{the Schrodinger wave equation}. It is desired to
estimate that law. To aid in its estimation, the dimensional projections
$a_{j}$ of \textbf{a} are repeatedly measured, a total of $k$ times, either in
identically prepared experiments, or $k$ times in the same experiment. (An
example of the latter is where the state \textbf{a} is that of spin, and it is
measured at $k$ positions in the system.)\ 

The measurements are the \textbf{y}$_{j}\equiv y_{ji},i=1,...,k.$These are
generally imperfect versions of $a_{j}$, obeying%
\begin{equation}
\text{\textbf{y}}_{j}\mathbf{\ }\equiv~a_{j}+\text{\textbf{x}}_{j}%
\text{\textbf{, }or }y_{ji}=a_{j}+x_{ji}\text{\textbf{, } with }0\leq
a_{j}\leq l\text{ and}~j=1,...,n,~i=1,...,k. \tag{1}%
\end{equation}
\ The unknown state \textbf{a} thereby lies within an $n$-dimensional 'box' of
length $l$. Also, the data \textbf{y}$_{j}$ suffer errors \textbf{x}%
$_{j}\equiv x_{ji},$random fluctuations on the continuum that are
\emph{characteristic of the physics of the particle}.\ It is assumed that
\emph{no additional fluctuations due to, say, noise of detection enter in}.
The point of the measurement experiment is to determine the physics of the
fluctuations \textbf{x}$_{j}$, which are considered to define the system of
state \textbf{a}. Thus there are, in total, $nk$ measurements of $n$ scalar
parameters $a_{j}.$

\subsubsection{Generalizing Hardy's axiom $N=\max$}

Let two states \textbf{a} of the system be \emph{distinguishable} if an
observer can tell which is present when viewing a single copy \textbf{y} of
the system. Thus, how many states \textbf{a} of the system are
distinguishable? Call this $N$. \ It is intuitive that the less random are the
measurement errors \textbf{x} the larger should be $N.$\ Hardy postulates,
using an idealized spin experiment [1] where in effect fluctuations
\textbf{x}$_{j}=0,$\ that in all physical cases, whether classical or quantum,%
\begin{equation}
N=\max. \tag{2}%
\end{equation}

Instead, in our scenario Eq. (1), the observables are data \textbf{y}$_{j}$
and these are \emph{generally imperfect }versions of $a_{j}$ due to
fluctuations \textbf{x}$_{j}.$\ We wish to define the physics of the
fluctuations \textbf{x}$_{j}.$\ Because of these fluctuations in the data,
system property (2) is not immediately apparent. Instead,\emph{ we regard it
as a corollary of the Hardy axioms}.

The value of $N$ can only be estimated, on the basis of the data obeying Eqs.
(1). Intuitively, $N$ \ should increase as $l/\delta x$ (quantified below)
where $\delta x$ is\ some measure of the uncertainty in\ fluctuations
\textbf{x} and $l$ is the total length of the box interval. Thus, because of
property (2) and the fixed nature of $l$, uncertainty $\delta x$ is assumed to
be minimal. Intuitively this suggests maximum information $I$ in the data, as
turns out to be the case at Eq. (15).

For simplicity of notation, denote all \textbf{y}$_{j}\mathbf{,}$%
\textbf{x}$_{j},a_{j},$ $j=1,...,n$ as \textbf{x},\textbf{y},\textbf{a}%
.\ \ Let the \textbf{y} occuring in the presence of the state \textbf{a}
\ obey a fixed likelihood density law $p($\textbf{y}$|$\textbf{a}$).$ This law
defines the physics of the data-forming system. The law is most conveniently
found within the context of \emph{estimating the unknown state a of the
system}.

\subsubsection{System fluctuations \textbf{x}$_{j}$ as additive noise}

The \textbf{x}$_{j}$ $\equiv x_{ji},$ $i=1,...,k$ are random, statistically
independent fluctuations from the unknown state values $a_{j},$ and are
assumed independent of values $a_{j}.$\ Then likelihood law $p($\textbf{y}%
$|$\textbf{a}$)=p_{X}($\textbf{x}$)$. As mentioned below Eq. (1), the physics
of the system lies in its fluctuations \textbf{x}$_{j}$. Thus each fluctuation
vector \textbf{x}$_{j}$ follows a generally different probability law $p_{j}%
($\textbf{x}$_{j}),$ $j=1,..,n.$ Observing data \textbf{y}$_{j}$ allows one to
estimate the system state \textbf{a}, but also (as will be seen) to estimate
the probabilities $p_{j}($\textbf{x}$_{j})$ defining these input channels.
\ This will be taken up in Sec. IV.\ Meanwhile, we return to the problem of
determining state \textbf{a}.

By Eq. (1), the $k$ numbers in each dimensional data vector \textbf{y}$_{j}$
suffer from\ $k$ corresponding random fluctuations \textbf{x}$_{j}=x_{ji},$
$i=1,...,k$ from the ideal state value $a_{j}.$ It is assumed, for simplicity,
that fluctuations \textbf{x}$_{j}$ are independent of the corresponding
parameter component $a_{j}.$ This describes a shift-invariant system. \ Of
course physical effects commonly obey shift invariance [2],[22]; this means,
e.g., that the physics of a given system does not change as it is moved from
one laboratory room to another. (Of course the potential source is assumed to
rigidly move with the rest of the system.) However, we first proceed as far as
possible (until Sec. IIIA) \emph{without} the assumption of shift invariance.

\subsubsection{Discrete, or continuous, nature of the \textbf{x}$_{j}$}

The state fluctuation values \textbf{x} are required by Hardy to be either (i)
finite in number or (ii) countably infinite in number. \ These correspond,
respectively, to either a \ (i) discrete coordinate system \textbf{x}%
$_{j}\equiv x_{ji}=i_{j}\Delta x$ with $i_{j}=0,1,...,l/\Delta x$, with
$\Delta x$\ the common finite spacing of positions in all dimensions; or a
\ (ii) continuous coordinate system of \textbf{x} on interval $(0,l)$, where
$\Delta x\rightarrow dx,$ a differential.\ Both alternatives (i) and (ii) are
included within the analysis by leaving unspecified the nature of the
expectations $<>$\ indicated in Eqs. (3) \emph{et seq}. defining the Fisher
information.\ They could be either integrals over probability densities or
sums over absolute probability laws. In the former case the $p_{j}($%
\textbf{x}$_{j})$ and $p($\textbf{y}$|$\textbf{a}$)$ are regarded as absolute
probabilities, in the latter probability \emph{densities}.

With choice (i) of $\Delta x$ finite, given also the finite size $l$ of the
cube the $l/\Delta x$ coordinates for each \textbf{x}$_{j}$ are finite in
number (choice (i) of Hardy). \ With choice (ii) there are an infinity of
fluctuation values \textbf{x}$_{j}$. \ However, as results below show, in
either case (i) or (ii) the resulting $N$ is generally finite, but can be
\emph{countably infinite} (choice (ii) of Hardy). \ This would arise when one
or more of the probability density functions $p_{j}($\textbf{x}$_{j})$ contain
regions of very high slope, as near edges or impulses. This causes the Fisher
information $I$ to approach infinity, implying that determinism is being
approached. \ 

\subsection{Hardy's requirement $N=N_{A}N_{B}$}

Consider subsystems $A$ and $B$ of the system, e.g. event states in two
different dimensions. This separability property $N=N_{A}N_{B}$ (see below Eq.
(6)) is Axiom 3 of Hardy. This Axiom and Eq. (2) are the usual properties of
accessible microstates in a statistical mechanical system at equilibrium,
although neither Hardy nor we assume an equilibrium state.

Thus\ Eq. (2) and Axiom 3 are assumed to hold for systems in general states of
non-equilibrium. In fact these are actually \emph{the only facets of Hardy's
approach} that are used to prove our required results that (i) the system's
Fisher information obeys $I=I_{\max}$, and that (ii) this holds for systems in
general states of non-equilibrium. See, e.g., past applications of $I=I_{\max
}$ to general non-equilibrium thermodynamics in [2],[4], and to living cells
in [11]-[14] (which must be in non-equilibrium states in order \emph{to be} alive).

By Hardy's approach, Eq. (2) and Axiom 3 also hold whether the system obeys
classical or quantum physics.

\section{Computing $N$}

The number $N$ of \emph{observably }distinguishable states of the system is
next shown to relate to its level of Fisher information.

\subsection{Fisher information}

Suppose that the system is 'coarse grained', defined as any perturbation of a
system that causes it to lose local structural detail. This takes place during
a small time interval $\Delta t$. The physical requirement is that this coarse
graining must, by perturbation of the state, cause it to generally lose
information, i.e. to obey $\delta I\leq0$ for $\Delta t>0.$ On this basis a
measure of 'order' was derived as well, in Ref. [15] for the case of a single
scalar dimension $x$, and in [16] for the most general case of a vector
coordinate \textbf{x }$=(x_{1},...,x_{n})$ of general dimension $n$. The order
measure has analogous properties to the measures of Kolmogoroff [26] and
Chaitin [27]. \ 

Coarse graining has the further property of demarking the transition from a
quantum to a classical universe [23]. Although Fisher's classic information
$I$ defined by Eq. (9) has been used to derive quantum mechanics [2] a version
of Fisher information specifically designed to apply to quantum scenarios is
being developed as well\ [24],[25].

\subsection{Fisher information matrix}

The principal aim of this paper is to show that extending Hardy's postulate
$N=\max$. to a scenario of noisy data, where distinguishable states are not
obvious, gives rise to a principle of maximization of the Fisher information.
\ First we relate $N$ to the Fisher information matrix. The latter is defined
[2],[22] as the $n\times n$ matrix $\left[  I\right]  $ of elements%

\begin{equation}
I_{ij}\equiv\left\langle \frac{\partial\ln p(\text{\textbf{y}%
$\vert$%
\textbf{a}})}{\partial a_{i}}\frac{\partial\ln p(\text{\textbf{y}%
$\vert$%
\textbf{a}})}{\partial a_{j}}\right\rangle ,~i,j=1,...,n\text{ independently.}
\tag{3}%
\end{equation}
In Eq. (3), the brackets $\left\langle {}\right\rangle $ denote an expectation
with respect to likelihood law $p($\textbf{y}%
$\vert$%
\textbf{a}$).$ \ Let the matrix $\left[  I\right]  $ have\emph{ an inverse}
$\left[  I\right]  ^{-1}.$ Denote the latter's $ij$th element as $I^{ij}$
(Note the superscripts.) The\ Cramer-Rao inequality states that the
mean-squared error $\varepsilon_{j}^{2}$ in determining\ $a_{j}$ has as its
lower bound the $j$th diagonal element of $\left[  I\right]  ^{-1},$%

\begin{equation}
\varepsilon_{j}^{2}\geq I_{{}}^{jj},\text{ or }1/\varepsilon_{j}\leq\left(
I_{{}}^{jj}\right)  ^{-1/2}. \tag{4}%
\end{equation}
In fact the lower bound errors $\varepsilon_{j}^{2}\equiv\varepsilon_{j\min
}^{2}=I^{jj}$ can be \emph{achieved} in systems whose laws $p($\textbf{y}%
$\vert$%
\textbf{a}$)$ obey an 'efficiency' condition [22].

\subsection{Alternative measures of error}

As will be seen, this specifically second-order measure of error
$\varepsilon_{j}\equiv<|$\textbf{x}$|^{2}>^{1/2}$ may be used to definine the
distinguishability of neighboring state values $a_{j}.$ However, certain
systems might instead require 3rd- or higher-order error measures.
\ Nevertheless, all\ such higher-order error measures are generally larger
than $\varepsilon_{j},$ as shown by\ \emph{Lyapunov's inequalities }[28]%
\begin{equation}
\varepsilon_{j}\equiv\sqrt{\varepsilon_{j}^{2}}\equiv<|\text{\textbf{x}}%
|_{j}^{2}>^{1/2}\leq\ <|\text{\textbf{x}}|_{j}^{3}>^{1/3}\leq
~<|\text{\textbf{x}}|_{j}^{4}>^{1/4}\leq\ ... \tag{5}%
\end{equation}
These will be shown at Eqs. (15) and (16) to give the same result $I=I_{\max}$
as does the use of 2nd-order measure $\varepsilon_{j}.$

\subsection{Number $N$ of distinguishable states}

\subsubsection{Quantum entanglement}

As a counter example, our scenario of multiple measurements $i=1,...,k$ of
$a$\ in each dimension $j$ is capable of encountering
\emph{quantum-entanglement}. \ This where measurement pairs $y_{ji}$ and
$y_{ji^{\prime}}$ are at least partially dependent for $i\neq i^{\prime}.$ The
\emph{intrinsic} fluctuations $x_{ji}$ and $x_{ji^{\prime}}$ then have the
same property of entanglement. This can be entanglement in states of position,
or momentum, spin, polarization, etc. \ An example is where, in the usual EPR
experiment, the daughter 1 particle spin component state $a_{fi}$ is measured
on the earth and the daughter 2 state $a_{ji^{\prime}}$ is simultaneously
measured on the moon. \ Their measurements $y_{ji},y_{ji^{\prime}}$ are found
to always point in opposite directions. This rules out contribution to $N_{j}$
of the two measurements pointing \emph{in the same} direction, thereby
reducing $N_{j}$ by 2 (states up-up and down-down). Thus, taking entanglement
cases into account over all dimensions $j$ can only reduce the number $N$ of
distinguishable states. \ In fact [29] "a set of entangled states can be
completely indistinguishable, that is, it is not possible to correctly
identify even one state with a non-vanishing probability; whereas a set
containing at least one product state is always conclusively distinguishable
(with non-zero probability)."

Ignoring, for the moment, the possibility of quantum entanglement (see above),
the maximum possible number of distinguishable component state positions
$a_{j}$ in projection $j$ obeys simply $N_{j}=$ $l/\varepsilon_{j}$ [30].
Therefore allowing for possible quantum entanglement can only reduce this
total, giving $N_{j}\leq$ $l/\varepsilon_{j}.$\ 

\subsubsection{Resulting upper bound to $N$}

Each of the $l/\varepsilon_{j}$\ states\ characterizing\ a dimension $j$ can
occur\ simultaneous with any one of the\ $l/\varepsilon_{j^{\prime}}$ states
of \emph{any other} dimension $j^{\prime}\neq j.$\ Therefore the total number
$N$ of distinguishable states \textbf{a}\ over \emph{all} dimensions obeys a
simple product%

\begin{equation}
N\leq\frac{l}{\varepsilon_{1}}\cdot\cdot\cdot\frac{l}{\varepsilon_{n}} \tag{6}%
\end{equation}
over the $n$ dimensions [29]. This product result also utilizes the Hardy
Axiom 3, that a composite system consisting of subsystems A and B satisfies
$N=N_{A}N_{B}.$ \ 

Note that the inequality sign in Eq.(6) is replaced by an equality sign for
classical systems (where entanglement does not occur and hence cannot reduce
the number of distinguishable states). \ But this will not affect the results
below, beginning with Eq. (7). The latter is already an inequality because of
inequality (4), which has nothing to do with entanglement. Hence the final
result $I=I_{\max}$ will hold for either classical or quantum systems.

Using inequalities (5) in Eq. (6) show that $N$ computed on the basis of 3rd-
or higher-order error measures is smaller than $N$ computed on the basis of
the 2nd-order error $\varepsilon_{j}.$ This will ultimately show (below Eq.
(16)) that the Fisher $I\ =\max$. even when such \emph{higher-order} error
measures are used.

\section{Maximizing $N$}

By Eq. (4), Eq. (6) is equivalent to%

\begin{equation}
N\equiv N_{m}\leq l^{n}\left(  I_{{}}^{11}\cdot\cdot\cdot I^{nn}\right)
^{-1/2}. \tag{7}%
\end{equation}
The subscript $m$ anticipates that $N_{m}$ will be maximized.

\subsection{Utilizing the independence\ of fluctuations \textbf{x}}

As mentioned in Sec. I A3, all fluctuations \textbf{x} are independent of one
another and of the states \textbf{a}, implying a shift-invariant system.

In this scenario the information obeys simple additivity [31]%

\begin{equation}
I=\sum\limits_{j=1}^{n}I_{j} \tag{8}%
\end{equation}
over the $n$ dimensions. Also, the matrix $[I]$ defined by Eq. (3) becomes
purely diagonal, with elements%

\begin{equation}
I_{jj}\equiv I_{j}\equiv\left\langle \left(  \frac{\partial\ln
p(\text{\textbf{y%
$\vert$%
a}})}{\partial a_{j}}\right)  ^{2}\right\rangle . \tag{9}%
\end{equation}
Finally, the inverse matrix $[I]^{-1}$ is likewise diagonal, with elements
$I^{jj}$ that are the reciprocals of corresponding elements $I_{j}$ of Eq. (9),%

\begin{equation}
I^{jj}\equiv I^{j}=1/I_{j}=\left\langle \left(  \frac{\partial\ln
p(\text{\textbf{y%
$\vert$%
a}})}{\partial a_{j}}\right)  ^{2}\right\rangle ^{-1} \tag{10}%
\end{equation}
Then Eqs. (7) and (9) give%

\begin{equation}
N_{m}\leq l^{n}\sqrt{I_{1}I_{2}\cdot\cdot\cdot I_{n}}=l^{n}\prod
\limits_{j,i}\left\langle \left(  \frac{\partial\ln p(\text{\textbf{y%
$\vert$%
a}})}{\partial a_{j}}\right)  ^{2}\right\rangle ^{1/2}. \tag{11}%
\end{equation}
Taking a logarithm gives%

\begin{equation}
\ln N_{m}\leq n\ln l+\frac{1}{2}\sum\limits_{j=1}^{n}\ln I_{j}^{{}}. \tag{12}%
\end{equation}

Axiom (2) states that the left-hand side of Eq. (12) must be maximized. But by
the nature of the inequality in (12) the right-hand side is even larger. Then,
how large can it be?$\ $

\subsection{\bigskip Use of Lagrange undetermined multipliers}

Since the individual $I_{j}$ in Eq. (12) must obey the constraint Eq. (8) they
are not all free to vary in attaining the value of $N_{m}.$ This may be
accommodated by amending requirement (12) via use of undetermined Lagrange multipliers,%

\begin{equation}
\ln N_{m}=n\ln l+\frac{1}{2}\sum\limits_{j=1}^{n}\ln I_{j}+\lambda\left(
\sum\limits_{j=1}^{n}I_{j}-I\right)  =\max. \tag{13}%
\end{equation}
This has a single undetermined multiplier $\lambda.$ \bigskip Differentiating
(13) with respect to any $I_{j}$ and setting the result equal to zero gives an
interim solution that $I_{j}=const.$ $\equiv I_{0}$ for all $j=1,...,n.$ Also,
$\lambda=-n/(2I)=const.$ since $I=const$. The size $I_{0}$ of the common
information value $I_{j}$\ is found by requiring it to obey constraint (8)
This gives the value $I_{0}=I/n=$ $I_{j}$ for all $j=1,...,n$. Using this in
Eq. (13) and taking the antilog gives a value of%

\begin{equation}
N\equiv N_{m}=\left(  \frac{l^{2}}{n}I\right)  ^{n/2} \tag{14}%
\end{equation}
Thus, since $N$ is maximized by axiom (2), likewise%

\begin{equation}
I=I_{\max}. \tag{15}%
\end{equation}
Thus an $n$-dimensional system with independent data will obey Hardy's axiom
(2) if its total information $I$ is maximized. \ Showing this was a major aim
of the paper.

Principle (15), supplemented by constraint information, has in the past been
used to derive laws of statistical mechanics [2],[3],[4],[17], biology
[2],[3],[5],[6],[11]-[14],[20] and economics [2][3],[18],[19].

\subsection{Use of 3rd or higher-order error measures}

We noted below Eq.(6) that an alternative error measure $<|$\textbf{x}%
$|_{j}^{3}>^{1/3},$ $<|$\textbf{x}$|_{j}^{4}>^{1/4}$or higher-order produces
\emph{smaller} values of $N$ than does the specifically \emph{2nd-order}
meaure $\varepsilon_{j}^{2}.$ Call these respective $N$ values $N_{m}^{(3)},$
$N_{m}^{(4)},...$ For example, consider a case $N_{m}^{(3)}.$\ \ Then by Eqs.
(5) and (6), $N_{m}^{(3)}\leq N_{m}.$ Next, by the Hardy axiom Eq. (2)
$N_{m}^{(3)}$is to be maximized. Then the problem Eq. (13) is replaced with a problem%

\begin{equation}
\max.=\ln N_{m}^{(3)}\leq\ln N_{m}=n\ln l+\frac{1}{2}\sum\limits_{j=1}^{n}\ln
I_{j}+\lambda\left(  \sum\limits_{j=1}^{n}I_{j}-I\right)  . \tag{16}%
\end{equation}
The outer equality is identical with the maximization problem Eq.(13). Hence
it has the same solutions (14), (15). \ Thus $I=I_{\max}$ results by use of
any error measure of order $2$ or higher.

\subsection{Verifying that the extremized $N$ is a maximum}

However, seeking the solution to Eq. (13) by variation of the $I_{j}$
guarantees that it achieves an extreme value in $\ln N_{m}$ , but not
necessarily its maximum one. The extremum could, e.g.,have instead been a
minimum. \ For $\ln N_{m}$ to be maximized its second derivative must obey
$(\partial^{2}(\ln N_{m})/\partial I_{j}^{2})<0.$The first derivative of (13)
gives $\partial(\ln N_{m})/\partial I_{j}^{{}}=1/(2I_{j}),$ so the second
derivative gives $-1/(2I_{j}^{2})<0,$ negative as required. Thus $N_{m}=\max.$
as required.

\section{Significance to past $I$-based derivations of physics}

As mentioned at the outset, in general the system can be coherent, obeying a
complex vector amplitude function $\Psi($\textbf{x}$)$ such that%

\begin{equation}
p(\text{\textbf{x}})=\mathbf{\Psi}^{\ast}(\text{\textbf{x}})\cdot\mathbf{\Psi
}(\text{\textbf{x}})=|\mathbf{\Psi}(\text{\textbf{x}})|^{2}. \tag{17}%
\end{equation}
The physics obeyed by many systems specified by either scalar $p($%
\textbf{x}$)$ or vector $\mathbf{\Psi}($\textbf{x}$)$ laws have been found
[2-20] by the use of relation\emph{ }(15) \emph{as a hypothesis}. Here, by
comparison, we have shown that (15) is generally valid, i.e. consistent with
Hardy's axioms. \ Thus it is no longer merely a hypothesis.

\subsection{Need for prior knowledge-based constraints}

But what value of $I$ will result from its maximization? Mathematically, the
maximum possible value of $I$ is infinity (by Eq. (9), occuring if edge- or
impulse-like details exist in $p,$ causing one or more slope values
$\partial\ln p/\partial a_{j}$ to approach infinity). \ But infinite values of
$I$ could, by the Cramer-Rao inequality (4) in an 'efficient' scenario (see
the sentence following (4)), allow zero error,\ i.e. perfect determinism, to
be present. This is contrary to the assumption that the system is statistical.
We of course wish to avoid such inconsistent uses of the principle $I=I_{\max
}$. To avoid the causative 'runaway' value for $I$ requires\emph{ prior
knowledge of a constraint on} $I$ that limits it to finite values. Such a one
was used in problem (13), where the $\lambda-$term was added on for precisely
that purpose. (As an actual application to biology, principle (15) was used to
derive the law of natural selection [20].) But does the constraint term have a
\emph{physical} meaning?

\subsection{Physical meaning of constraint term}

Thus, the principle we now seek that avoids predictions of infinite slope, and
determinism, must take the form%

\begin{equation}
I-J=extremum,\text{ where }J=I_{\max} \tag{18}%
\end{equation}
and \ the latter is of \emph{finite} value. This is called the principle of
'extreme physical information' or EPI. The extremum\ achieved is nearly always
a minimum. The solution $p($\textbf{y%
$\vert$%
a}$)=p($\textbf{x}$)$ or\ $\Psi($\textbf{y}$|$\textbf{x}$)=\Psi($\textbf{x}$)$
to (18) defines the physics obeyed by the system [2]-[5], [7] ,[10], [15],
[17], [19], [32]. These probability- and amplitude laws are the mathematical
solutions to the 2nd-order Euler-Lagrange differential equation solutions to
principle (18). Examples are the Klein-Gordon or Schrodinger wave equation
[2], [32].\ 

In principle (18), $J$ is the \emph{physical} representation, or source, of
the \emph{observed} information $I$. This can take the form of squared
particle momentum, or charge and/or energy flux, depending upon application.
Also, since $J$ is the source of $I,$ it represents the \emph{maximum possible
value} $I_{\max}$ of $I$. This works mathematically because the extremum in
Eq. (18) is nearly always a minimum [2]. So with $I-J=\min.,$ necessarily
$I\rightarrow J$ $=I_{\max}$ in value. Thus, $I$ is forced \emph{toward its
maximum value }$I_{\max},$ as required by principle (15). In particular, any
solution for $I$ attains some\emph{ fraction} of $I_{\max}$ between $1/2$ and
$1$, depending on application (e.g. value $1/2$ in classical physics and $1$
in quantum physics).\ Thus, mathematically, the finite nature of source term
$J$ acts as a constraint that limits the maximized value of $I$. Thus, it
plays the analogous role to the $\lambda-$dependent term in principle (16). \ 

\ And, as we saw, principle (18) holds for a classical or quantum system in
any state of non-equilibrium.

\section{Discussion}

\ In a quantum system, the amount of information encoded is limited by the
dimension of its Hilbert space, i.e., its number of perfectly distinguishable
quantum states \textbf{a}. In past work [33], principle (2) that $N=\max.$ was
shown to hold, in particular, for a binary quantum system of unit vector
states, provided the \emph{Shannon} mutual information between the (random)
quantum state and the results of its measurements is maximized. \ Thus, the
Shannon replaced our use here of the Fisher information. \ This Shannon result
was derived in cases where the angle between the two states in a
two-dimensional vector basis space is distributed with uniform randomness over
its allowable range. \ 

The issue of \emph{why} the concept of $N$, the maximum number of
distinguishable states, should be used as a founding axiom of physics is of
interest. It actually ties in with Godel's two incompleteness theorems.
According to these any finite axiomatic theory must be incomplete in scope
(i.e., admit of undecidable theorems). But it must at least be
self-\emph{consistent}. \ In fact the concept of a quantum holographic
universe [33] is\emph{ consistent with} a maximum $N$ (of size $N\sim
\exp(10^{10})$). Likewise, in statistical mechanics, the maximum
\emph{consistent} event \textbf{a} is the one that can occur in the\emph{
maximum} number of ways $N$ [35].

There are a fixed total $nk$ of outcomes resulting from the measurements
described in Eq. (1). \ Then the result (15) that $I=I_{\max}$ also means that
the amount of information per measurement obeys $I/nk=\max$ \ In turn, this
may be interpreted as\ meaning that any observed system is "maximally free"
[36], in the sense that a specification of its statistical state \textbf{a}
that is sufficient for predicting the probabilities of outcome of
\emph{future} measurements \textbf{y} will require the \emph{maximum amount of
information} $I$ per experimental outcome.\ \ In this sense the experiment is
minimally predictive, thereby exhibiting "free will." \ Moreover, as we saw,
$I=I_{\max}$ holds for classical or quantum systems. This raises the
possibility that the molecular neurons of the brain, regardless of whether
they obey classical or quantum [37] physics, exhibit free will. This might
translate, in turn, into \emph{conscious} free will [38]. However the evidence
for this is not yet compelling.

Our principal use of Hardy's axioms was the postulate\ Eq. (2), \emph{which
holds (a) whether the system is classical or quantum, and also (b) whether it
is in any state of equilibrium or non-equilibrium.} \ Property (a) must also
relate to the fact that $I$ is a limiting form [2], [39] of the density
function $\rho($\textbf{x}$,$\textbf{x}$^{\prime})$ of the system, which
represents the system in the presence of only partial knowledge of its phase
properties. This describes a mesoscopic system, one partway between being
classical and quantum. Thus, our main results Eqs. (15) and (18), which state
that $I=I_{\max}$, likewise hold under these general conditions (a) and (b).
Then these justify past uses of EPI principle (18) to derive [2-20] diverse
laws of statistical physics, including those of cell biology, whose living
cells \emph{must be in non-equilibrium states}. \ \ \ 

In summary, the EPI\ principle (18) and information maximization principle
(15) are no longer merely convenient hypotheses for deriving statistical laws
of physics. Rather, they are \emph{a priori} known to be valid for all current
physics on its most basic level, that of its underpinnings in Hardy's
mathematical axioms. This fills an important gap in all past applications
[2-20] of the principle of maximum $I$\ to physics, biology and econophysics,
which \emph{assumed}, as a working hypothesis, that $I=I_{\max}$. \ It also
justifies future uses of principles (15) and (18).

\section{Acknowledgment}

The authors acknowledge support from the National Cancer Institute under grant
1U54CA143970-01. One of us (BRF) thanks Prof. J.O. Kessler for suggesting that
a principle of maximum Fisher information might conceivably be implied by L.
Hardy's axioms.

\textbf{REFERENCES}

1. L. Hardy, "Quantum theory from five reasonable axioms," manuscript
quant-ph/0101012 (2003); also, L. Hardy, "Probability theories in general and
quantum theory in particular," Studies in History and Philosophy of Modern
Physics \textbf{34}, 381-393 (2003)

2. B.R. Frieden, S\emph{cience from Fisher Information, 2nd ed.} (Cambridge
University Press, United Kingdom, 2004)

3. B.R. Frieden and R.A. Gatenby, eds., \emph{Exploratory Data Analysis Using
Fisher Information} (Springer, London, 2007)

4. B.R.Frieden, A. Plastino, A R Plastino, B H Soffer,\ "Schrodinger link
between nonequilibrium thermodynamics and Fisher information,"\ Phys. Rev. E
\textbf{66}, 0461328 (2002)

5. B.R. Frieden and R.A. Gatenby, "Power laws of complex systems from extreme
physical information,"\ Phys. Rev. E \textbf{72}, 036101 (2005)

6. R.A. Gatenby and B.R. Frieden, "Inducing catastrophe in malignant
growth,"\ J. Math. Med. Biol. \textbf{25}, 267-283 (2008)

7. B.R. Frieden and B.H. Soffer, "De Broglie's wave hypothesis from Fisher
information,"\ Physica A \textbf{388}, 1315-1330 (2009)

8. B.R. Frieden, A. Plastino and A.R. Plastino, "Effect upon universal order
of Hubble expansion,"\ Physica A doi: 10.1016/j.physa.2011.08.005 (2011)

9. B.R. Frieden and M. Petri, "Motion-dependent levels of order in a
relativistic universe,"\ Phys. Rev. E \textbf{86}, 032102 (2012)

10. R. Carroll, \emph{On the Quantum Potential} (Arima, Suffolk, United
Kingdom, 2007)

11. R.A. Gatenby and B.R. Frieden, \textquotedblleft Coulomb Interactions
between Cytoplasmic Electric Fields and Phosphorylated Messenger Proteins
Optimize Information Flow in Cells,\textquotedblright\ PLoS ONE 5(8): e12084.
doi:10.1371/journal.pone (2010)

12. B.R. Frieden and R.A. Gatenby, \textquotedblleft Information Dynamics in
Living Systems: Prokaryotes, Eukaryotes, and Cancer,\textquotedblright\ PLoS
ONE 6(7): e22085. doi:10.1371/journal.pone.0022085 (2011)

13. R.A. Gatenby and B.R. Frieden, \textquotedblleft Cell Development Pathways
Follow from a Principle of Extreme Fisher Information,\textquotedblright\ J.
Physic Chem Biophysic 2011, 1:1 http://dx.doi.org/10.4172/2161-0398.1000102

14. B. R. Frieden and R. A. Gatenby (invited paper) \textquotedblleft Cell
development obeys maximum Fisher information,\textquotedblright\ Frontiers in
Bioscience E5, 1017-1032 (2013)

15. B.R. Frieden and R.J. Hawkins, "Quantifying system order for full and
partial coarse graining," Phys. Rev. E \textbf{82}, 066117, 1-8 (2010)

16. B.R. Frieden and R.A. Gatenby, "Order in a multiply-dimensioned
system,"\ Phys. Rev. E \textbf{84}, 011128, 1-9 (2011)

17. B.R. Frieden, A. Plastino, A.R. Plastino and B.H. Soffer, "Fisher-based
thermodynamics:\ Its Legendre transform and concavity properties,"\ Phys. Rev.
E \textbf{60}, 48053 (1999)

18. R.J. Hawkins and B.R. Frieden, "Fisher information and equlibrium
distributions in econophysics,"\ Phys. Lett. A \textbf{322}, 126 (2004)

19. R.J. Hawkins, M. Aoki and B.R. Frieden, "Asymmetric information and
macroeconomic dynamics,"\ Physica A \textbf{389}, 3565-3571 (2010)

20. S.A. Frank, "Natural selection maximizes Fisher information," J. of
Evolutionary Biology \textbf{22}, 231-4 (2009)

21. R.A. Fisher, "On the mathematical foundations of theoretical statistics,"
Phil. Trans. R. Soc. Lond. \textbf{222}, 309 (1922)

22. B.R. Frieden, \emph{Probability, Statistical Optics and Data Testing, 3rd.
ed }(Springer-Verlag, Berlin, 2001); H.L. Van Trees, \emph{Detection,
Estimation and Modulation Theory, Part 1} (Wiley, N.Y., 1968), pp. 79-81;

23. J. J. Halliwell, "How the quantum universe became classical," Contemp.
Phys. 46, 93 (2005)

24. S.-L. Luo, "Wigner-Yanase skew information vs. quantum Fisher
information," Proc. American Mathematical Soc. 132, 885-890 (2003); also
"Logarithm versus square root: comparing quantum Fisher information," Commun.
Theor. Phys. 47, 597 (2007)

25. P. Gibilisco, D. Imparato and T. Isola, "A Robertson-type uncertainty
principle and quantum Fisher information," Linear Algebra and its Applications
\textbf{428},1706-1724 (2008); also P. Gibilisco, F. Hiai, and D. Petz,
"Quantum Covariance, Quantum Fisher Information, and the Uncertainty
Relations," IEEE Trans. Inf. Theory \textbf{55}, 439 (2009).

26. A.N. Kolmogoroff, "Three Approaches to the Quantitative Definition of
Information,"Problems Inform. Transmission 1 (1): 1-7(1965)

27. G.J. Chaitin, "On the simplicity and speed of programs for computing
infinite sets of natural numbers," JACM \textbf{16}, 407-422 (1969)

28. V.K. Rohatgi, \emph{An Introduction to Probability Theory and Mathematical
Statistics} (Wiley, New York, 1976)

29. S. Bandyopadhyayar,\ "Entanglement and perfect LOCC discrimination of a
class of multiqubit states," arXiv:0907.1754v2 [quant-ph] 27 Jan 2010.

30. M.R. Foreman and P. Torok, "Information and resolution in electromagnetic
optical systems," Phys. Rev. A \textbf{82}, 043835 (2010). The number of
distinguishable states in, specifically, polarization space is defined to be
the ratio of the volume of uncertainty in polarization before a measurement to
the volume of uncertainty after a measurement. In our problem, in a given
dimension $j$\ the former uncertainty is $l$, the latter $\epsilon_{j}.$

31. A. Engel and C. Van den Broeck, \emph{Statistical Mechanics of Learning}
(Cambridge University Press, U.K., 2001)

32. B.R. Frieden and B.H. Soffer, "Lagrangians of physics and the game of
Fisher-information transfer,"\ Phys. Rev. E \textbf{52}, 2274-2286 (1995)

33. L.B. Levitin, T. Toffoli, Z.D. Walton, "Information and Distinguishability
of Ensembles of Identical Quantum States," arXiv
$>$
quant-ph/0112075 IQSA 2001

34. C.J. Hogan, "Holographid discreteness of inflationary perturbations,"
arXiv:astro-ph/0201020v2 27 May 2002

35. F. Reif, \emph{Fundamentals of Statistical and Thermal Physics}
(McGraw-Hill, N.Y., 1965)

36. L. Smolin, "Precedence and freedom in quantum physics," arXiv:1205.3707
[quant-ph] May 17, 2012

37. W.R. Loewenstein, \emph{Physics in Mind: A Quantum View of the Brain}
(Basic Books, N.Y., 2013)

38. S.R. Hameroff and R. Penrose, \textquotedblleft Conscious events as
orchestrated spacetime selections,\textquotedblright\ Journal of Consciousness
Studies \textbf{3}(1): 36-53 (1996)

39. B.R. Frieden, \textquotedblleft Relations between parameters of a
decoherent system and Fisher information,\textquotedblright\ Phys. Rev. A
\textbf{66}, 022107 (2002)

\end{document}